# Titan's interaction with the supersonic solar wind

**Authors:** C. Bertucci[1*], D. C. Hamilton[2], W. S. Kurth[3], G. Hospodarsky[3], D. Mitchell[4], N. Sergis[5], N.J.T. Edberg[6], M. K. Dougherty[7].

**Affiliations:**

[1]Instituto de Astronomía y Física del Espacio (CONICET/UBA), Ciudad Universitaria, Buenos Aires, Argentina.

[2]Physics Department, University of Maryland, College Park, MD, United States.

[3]Department of Physics and Astronomy, University of Iowa, Iowa City, United States.

[4]Applied Physics Laboratory, Johns Hopkins University, Laurel, MD, United States

[5]Office for Space Research, Academy of Athens

[6]Swedish Institute of Space Physics, Uppsala, Sweden

[7]Blackett Laboratory, Imperial College London, London, United Kingdom,

*Correspondence to: cbertucci@iafe.uba.ar.

Keywords: Titan, Solar wind, Solar wind interaction with unmagnetized objects

**Abstract**

After 9 years in the Saturn system, the Cassini spacecraft finally observed Titan in the supersonic solar wind. These unique observations reveal that Titan's interaction with the solar wind is in many ways similar to un-magnetized planets Mars and Venus in spite of the differences in the properties of the solar plasma in the outer solar system. In particular, Cassini detected a collisionless, supercritical bow shock and a well-defined induced magnetosphere filled with mass-loaded interplanetary magnetic field lines, which drape around Titan's ionosphere. Although the flyby altitude may not allow the detection of an ionopause, Cassini reports enhancements of plasma density compatible with plasma clouds or streamers in the flanks of its induced magnetosphere or due to an expansion of the induced magnetosphere.

Because of the upstream conditions, these observations are also relevant for unmagnetized bodies in the outer solar system such as Pluto, where kinetic processes are expected to dominate.

## 1. Introduction

The absence of an intrinsic magnetic field at Titan (Ness et al., 1982) results in a direct interaction of the plasma environment with its ionized atmosphere. In the absence of collisions, this interaction consists of the electromagnetic coupling between the charged particles from Titan's ionosphere and its neutral corona, and the external plasma flow around the moon. The external flow approaching Titan progressively slows down as it is loaded with ions from its extended exosphere, but also as the moon´s ionosphere acts as a conducting obstacle. As a result, the magnetic field lines frozen into the external flow pile-up near the sub-flow point and stretch along the direction of the flow in the flanks and the downstream sector, defining an induced magnetosphere and magnetotail where the cold and dense plasma from Titan dominates (Neubauer et al., 2006). This interaction has been shown to lead to the removal of ionized atmospheric constituents (e.g. Coates, et al. 2012, Romanelli et al., Submitted to JGR, 2014), which may have important implications on the evolution of Titan's atmosphere.

Titan orbits Saturn at an average distance of 20.2 $R_S$ ($R_S$ = Saturn Radius = 60268 km ) near the Kronian equatorial plane. As a result, it spends most of its time in Saturn's partially co-rotating, magnetospheric flow (Thomsen et al., 2013). This flow transports Saturn's magnetic field, which encounters Titan at speeds of ~100 km/s and generates an interaction which is subAlfvénic and subsonic (Arridge et al., 2011). Titan has also been observed in Saturn's magnetosheath during periods when an increase in the solar wind pressure $P_{SW}$ compressed Saturn's magnetopause inside Titan's orbital radius (Bertucci et al., 2008), but no supersonic solar wind interaction has been observed until now.

In situ observations around Venus and Mars paved the way for the interpretation of Cassini measurements around Titan (Bertucci et al., 2011). However, all previous flybys had revealed strong differences with these planets, as Titan's magnetosphere displayed a far more complex structure of the magnetic field and plasma. These differences were attributed to Titan's richer atmospheric chemistry and variable plasma environment (e.g. Ulusen et al., 2012).

After 9 years in the Saturn system, the Cassini spacecraft observed the plasma environment of Titan in the supersonic solar wind for the first time. In this work we analyze the plasma boundaries and regions characterizing this interaction and discuss the physical processes behind them. In particular we discuss the importance of kinetic effects for induced magnetospheres in the outer solar system.

## 2. Cassini in situ Observations

The Cassini's T96 close encounter with Titan occurred on 1 December, 2013 at a local time of 12.52 hours with respect to Saturn. The time of closest approach was 00:41:18 UT. Figure 1 displays Cassini's trajectory in Titan-centered solar wind interaction (TSWIS) coordinates. In this coordinate system, the X-axis points anti-sunward, the Y-axis points in the direction of Saturn's orbital motion, and the Z-axis points north of Saturn's orbital plane (SOP). The spacecraft is in a highly inclined orbit and travels southward and tailward during the encounter. As a result, it explores the subsolar region on the inbound leg, followed by the southern terminator sector on the outbound leg.

The measurements of Cassini Magnetometer (MAG) (Dougherty et al., 2004), Magnetospheric Imaging Instrument/Charge Energy Mass Spectrometer (MIMI/CHEMS) (Krimigis, et al., 2004), and Radio and Plasma Wave Science instrument (RPWS) (Gurnett, et al., 2004) provide a detailed description of the plasma properties in the vicinity of Titan (Figure 2). T96 took place in the middle of a series of crossings of Saturn's bow shock. Cassini detects at least 16 crossings of Saturn's bow shock between November 28 and December 1 at kronocentric distances around 20.2 $R_S$ (not shown). In particular, T96 takes place in the supersonic solar wind between kronian bow shock crossings occurring at 18:18 on 30 November (outbound) and at 02:39 on 1 December (inbound). During this interval, the extension of the background interplanetary magnetic field (IMF) intersects Saturn's bow shock average location (Masters, et al., 2008), suggesting that the encounter occurs within Saturn's foreshock. The appearance of Langmuir waves in the range 6-7 kHz (also called electron plasma oscillations) both prior to 00:24 and after ~ 01:45 on the day of the encounter, (both before and after the Titan flyby) provide additional evidence of Titan's position in the supersonic solar wind, as these emissions are a common feature of the solar wind upstream of planetary bow shocks.

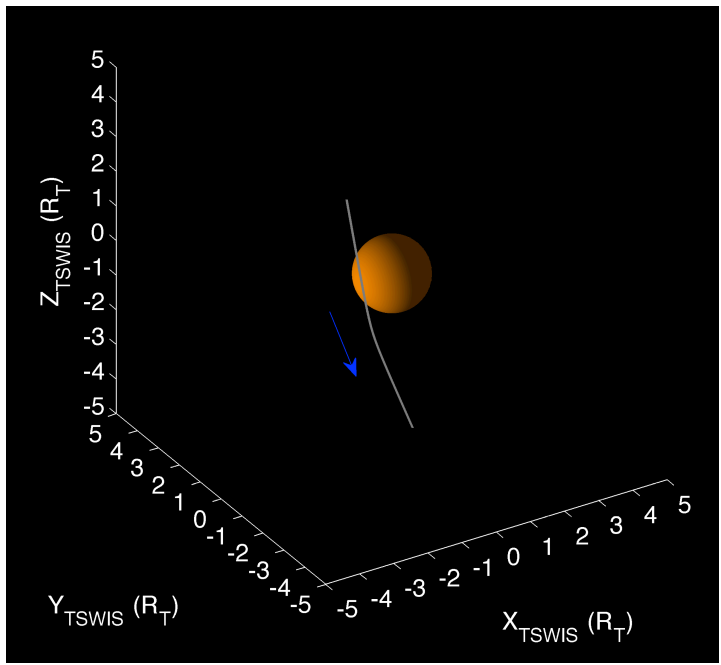

**Figure 1**. Trajectory of the Cassini Spacecraft in the vicinity of Titan in TSWIS coordinates,. The blue arrow indicates the sense of the spacecraft's path.

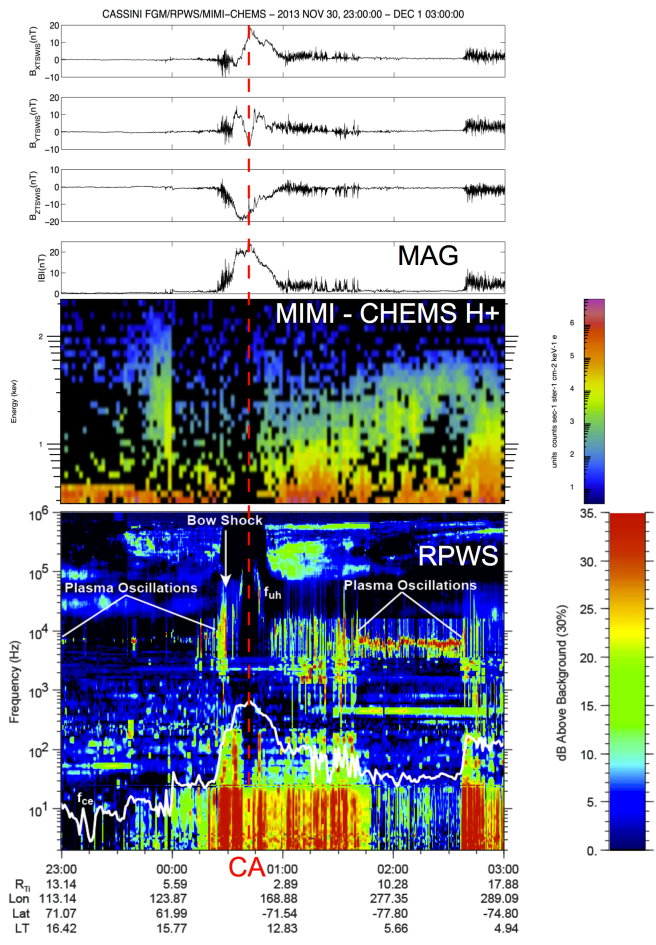

**Figure 2**. Magnetic field data in TSWIS coordinates from MAG, $H^+$ fluxes from MIMI CHEMS, and electric field spectrogram and local electron cyclotron frequency ($f_{ce}$) from RPWS during the T96 flyby. Closest approach crossings of Titan's inbound bow shock are indicated. Cassini's distance to Titan is indicated beneath the plots.

As Cassini approaches Titan, strong intensities of three species of pickup ions ($H^+$, $H_2^+$, and $O^+$) start to be observed by MIMI/CHEMS at about 21:50 on 30 November (even earlier for $H^+$) at 23 $R_T$ from the moon. The cutoff energies of all three species are about 4.3 keV/nuc corresponding to a solar wind speed of ~460 km/s assuming that most of their kinetic energy is contained in the direction perpendicular to the IMF. The latter statement is supported by the fact that from 21:00 on 30 November until 00:24 on 1 December, the cone angle of the IMF remains close to 90°. The observed pick-up ions may originate from Saturn's extensive neutral cloud (Sergis et al., 2013), although part of them may also come from Titan.

Around 00:00 UT on 1 December, Cassini crosses a shock front (hereafter SF). The detection of pick up ions with energies as high as 200 keV just before the SF is a result of their interaction with the shock potential and may be part of its foot. The SF is characterized by a factor-of-two enhancement and a rotation of the IMF, followed by a deceleration of the solar wind speed. The last pickup ion signature detected by CHEMS before T96 (~ 00:20 UT) yields a solar wind velocity of ~ 300 km/s and a dynamic pressure of ~0.09 nPa (the averaged solar wind density is 0.6 cm$^{-3}$ according to RPWS). In spite of these changes, the solar wind plasma from 00:00 until 00:24 remains supersonic in average: Alfven and magnetosonic Mach numbers (assuming a polytropic index $\gamma$ = 5/3 and an electron temperature of 1 eV) are 11 and 10 respectively. In the outbound leg pick up ion signatures are less evident, precluding a reliable estimate of the solar wind speed. Table 1 summarizes averaged inbound and outbound upstream parameters deduced from Cassini measurements.

Table 1: Average upstream parameters during T96 (times correspond to 1 December 2013).

|  | Inbound (00:00 - 00:24) | Outbound (01:45 - 02:10) |
|---|---|---|
| Electron density (cm$^{-3}$) | 0.6 | 0.50 |
| Solar wind speed (km s$^{-1}$) | 360 | - |
| Magnetic field strength (nT) | 0.98 | 1.08 |
| $M_A$ | 11 | - |
| $M_{MS}$* | 12* | - |
| Larmor Radius (km)* | 4396 | - |
| Ion inertial length (km) | 290 | 328 |

* Assuming electron and ion temperatures of 1eV.

The presence of Titan in the solar wind is noticeable in the Cassini data from the peaks in the magnetic field strength, plasma density, and the "bite-out" in the energetic H$^+$ around CA. The spacecraft encounters the moon's bow shock (BS) at a distance of 2.69 R$_T$ at a solar zenith angle (SZA) of 45.9° (00:24 UT) where Cassini observes a deceleration of the solar wind plasma, a sudden increase in magnetic field strength and fluctuations in both the magnetic field and plasma wave spectrum typical of a planetary magnetosheath (Figure 3). Coplanarity analysis yields an IMF-shock normal angle θ$_{Bn}$ = 60.2°.

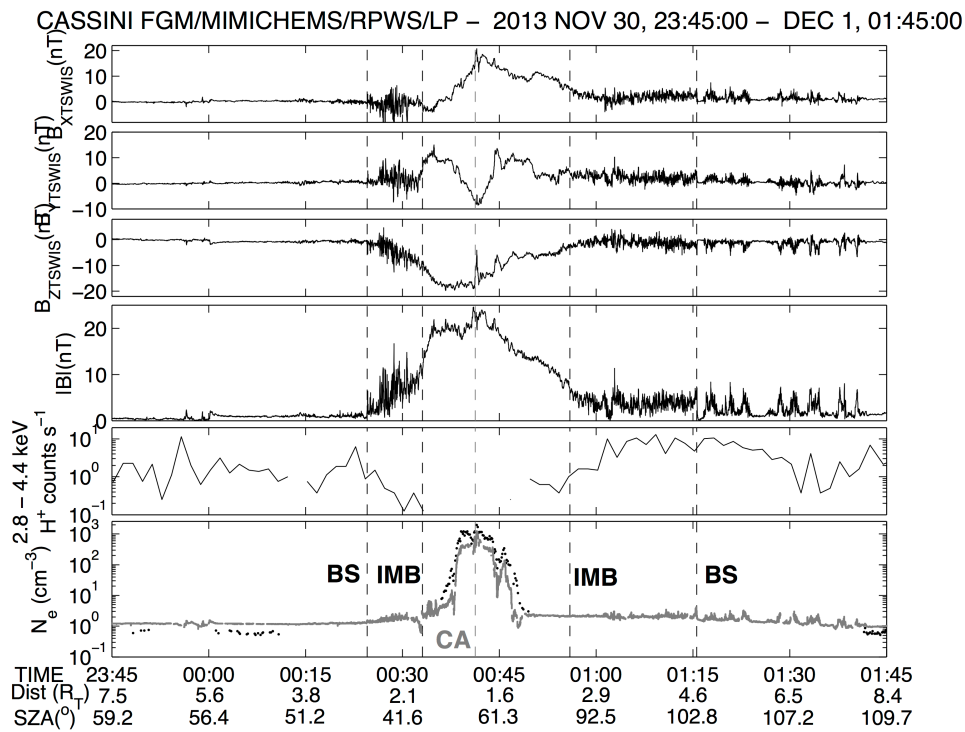

Figure 3. Magnetic field data in TSWIS coordinates (from MAG), H+ counts per second in the range 2.8-4.4 keV from MIMI CHEMS and total electron density (N$_e$) from RPWS (dots) and LP (grey curve) during the T96 flyby. CA, and crossings of Titan's induced magnetospheric boundary (IMB) and bow shock (BS) are indicated. Cassini's altitude above Titan (in Titan radii) and Solar Zenith Angle (SZA) are indicated.

Turbulent magnetic and electric field fluctuations observed by MAG and RPWS are observed from the bow shock and throughout the magnetosheath region. The magnetosheath fluctuations stop at the induced magnetospheric boundary (IMB), which displays remarkable similarities with that observed around active comets, Venus, and Mars (Bertucci et al., 2011). As the spacecraft crosses the IMB at 1.88 $R_T$ and at a SZA of 40.1° (00:33 UT), the magnetic field rotates and it becomes stronger. Minimum variance analysis (Sonnerup and Scheible, 1998) yields a well-defined boundary, with an intermediate-to-minimum eigenvalue ratio of 16.7 and a normal almost perpendicular to the mean magnetic field ($\theta_{Bn}$ = 80.8 ± 1.1°). A few minutes later, the electron density - determined from the upper hybrid resonance emission and proportional to the total plasma density - also increases. Inside the IMB, the magnetic field and the plasma density remain high.

The draping of the southward interplanetary magnetic field as the spacecraft transits the induced magnetosphere is clearly seen when an initially dominant $B_{ZTSWIS}$ is followed by an increase in $B_{XTSWIS}$ as the spacecraft travels south from Titan.

At CA, Cassini is marginally below Titan's nominal exobase (Garnier et al., 2007) and above the average ionospheric electron density peak altitude. The electron densities measured by Cassini RPWS and the Langmuir Probe (LP) at 00:41:24 are respectively 1245 and 1985 cm$^{-3}$. These estimates are consistent with previous measurements at similar altitudes and SZA (Agren et al., 2009; Edberg et al., 2010), suggesting that the upstream pressure may not influence ionospheric densities at those altitudes. At the same time, the plasma density around CA might not be as high as to suggest that magnetic diffusion dominates. It is precisely the absence of an appreciable decrease in the magnetic field around CA suggests that the convection of the magnetic field may still dominate at those altitudes (Cravens et al., 2010).

The IMB and BS are also detected in the outbound leg of the T96 flyby, suggesting that they are permanent features in Titan's supersonic interaction. Nevertheless, the inclination of Cassini's trajectory with respect to the Sun-Titan line gives rise to an expected inbound/outbound asymmetry in the location of these plasma features with respect to the nominal direction of the solar wind. In particular, a slightly elongated mirror image is observed in the magnetic field strength with respect to the inbound leg with the IMB and BS crossings occurring tailward, at 2.45 $R_T$ (00:56 UT), and 4.72 $R_T$ (01:16 UT) respectively. Additional asymmetries include: a) the presence of energetic (a few tens of keV) ions throughout the shocked and un-shocked solar wind and b) a series of high-amplitude fluctuations in the magnetic field and plasma wave spectrum upstream from Titan's outbound BS (between 01:17 and 01:42 UT).

## 3. Discussion

This is the first time that Titan is found in the supersonic solar wind. The plasma environment near T96 is, however, substantially variable and therefore, a careful analysis of the structures detected by Cassini in the surroundings of Titan is needed in order to correctly characterize the flow with which the moon was directly interacting at the time of the encounter. As described above, the shock front (SF) around 00:00 on 30 November is responsible for drastic changes in the solar wind conditions. The origin of this feature remains as an elusive issue, mainly because of the absence of CAPS measurements, and the lack of plasma measurements upstream from the spacecraft. However, important properties of the SF can be inferred from Cassini magnetic field and superthermal plasma measurements. First, the orientation of the IMF on either side of the SF intersects Saturn's shock surface model, implying that the SF occurs inside Saturn's foreshock. Second, if Titan is the obstacle generating the SF, the solar wind downstream from the SF to be subsonic, as the spacecraft frame is only moving at a few km s$^{-1}$ with respect to the moon. However, the MIMI CHEMS pickup ion measurements suggest that solar wind remains supersonic between 00:00 and 00:24. Third, it is unlikely that the SF be the inbound counterpart of a Titan-centered outer bow shock since no similar feature is observed after the encounter. Still, the role of Titan in the formation of the SF should be investigated in future works.

In contrast to the SF, the fast shock signatures (BS) at 00:24 and 01:16 on 1 December are clearly associated with Titan. Furthermore, the solar wind flow ahead of Titan's shock is supercritical (see Table 1), implying that part of the solar wind ions are expected to be reflected back upstream as dissipation is insufficient to account for the required retardation, thermalization and entropy increase (Treumann, 2009). In the inbound leg, the increase of wave activity around 00:15 may be indicative of an inbound Titan shock foot of the order of a local proton gyroradius in thickness. In addition, Titan's inbound shock does not display a clear ramp but is made of a series of magnetic field pulsations with increasing amplitude that precludes the identification of the typical substructures of supercritical shocks. Similar crossings have been observed at Mars (Bertucci et al, 2005a) and attributed to finite-gyroradius effects and exospheric ion pickup. In the outbound leg, the magnetic pulsations detected by Cassini after the outbound bow shock crossing are typical of planetary foreshocks (Schwartz et al., 1992). These pulsations likely arise from the nonlinear evolution of ultra low frequency (ULF) plasma waves originating from solar wind ions reflected at Titan's shock. These waves are expected to propagate towards the Sun within the moon's foreshock at speeds smaller than the solar wind's. As these waves grow in amplitude and become nonlinear, they would be carried downstream against Titan's shock contributing to its reformation.

The extent of Titan's dayside magnetosheath region is comparable to the upstream proton gyroradius and, therefore, it may be very small to allow efficient plasma thermalization. However, the increase in the magnetic field strength by a factor of nearly 10 within the magnetosheath due to both the solar wind mass loading by planetary ions and the presence of an ionospheric obstacle significantly reduces the gyroradii of the incoming particles in the lower magnetosheath contributing to a more efficient dissipation.

The inbound IMB thickness along the boundary normal (~ 170 km) is of the order of the local ion inertial length considering a local ion density of 2 cm$^{-3}$ (~160 km) supporting the idea that Hall fields play an important role in its formation (Bertucci et al., 2005b).

The size of Titan's induced magnetosphere is another element in common with Mars and Venus. Cassini measurements show that the sizes of Titan's bow shock and IMB in terms of the moon's radius are comparable to those reported at those unmagnetized planets (Zhang et al., 2008; Edberg et al., 2008). However, Titan's induced magnetosphere is significantly smaller when measured in terms of the gyroradii of locally picked up particles by the solar wind. Similar effects are expected at other unmagnetized outer solar system objects such as Pluto (Delamere & Bagenal, 2004).

Although T96 occurs 6 hours after Cassini crossed Saturn´s bow shock, the passage of the SF suggests that at the time of CA, Titan's interaction might not have occurred under steady state conditions. The magnetic pressure in the barrier $P_{Mbarrier} = \frac{B_{barrier}^2}{2\mu_0}$ reaches a maximum of 0.24 nPa around 00:40:48, when Cassini is located at 49.3° SZA. If pressure balance is expected, the expression: $k\rho_{SW}u_{SW}^2 \cos(SZA)^2 = P_{Mbarrier}$, where k = 0.88 (Dubinin et al., 2006) suggests a solar wind dynamic pressure $P_{SW} = \rho_{SW}u_{SW}^2$ of 0.64 nPa which is significantly greater than the one measured by Cassini before the inbound BS. A possible interpretation is that the IMF flux accumulated in the barrier may have been acquired from the accretion of flux tubes convected by the faster solar wind observed by Cassini before 00:00 on 1 December, suggesting that the same fossilization of upstream fields reported for T32 (Bertucci et al., 2008) may also have occurred during T96. This means that during the encounter, Titan's induced magnetosphere may have been expanding as it entered a low pressure solar wind sector following the passage of the SF. The local peak in electron density observed around 00:45 (~ 230 cm$^{-3}$) may be either a signature of an expanding induced magnetosphere as the spacecraft heads back into the solar wind, or an evidence of plasma cloud or streamer.

The weak solar wind pressure observed after 00:00 has also consequences on the dynamics of Saturn's shock, which is expected to expand as well. In agreement with this, an expanding kronian shock passes by Cassini at 02:39 UTC, well after the T96 flyby.

According to a previous work (Garnier et al., 2007), it is expected that at altitudes above Titan's bow shock H$_2$ will be the most important exospheric species, followed by H. However, apart from the foreshock-related pulsations, exospheric pre-shock wave activity is virtually absent at the proton cyclotron frequency of H$^+$ and H$_2^+$. It is then possible that although Titan's exosphere is detected farther away from its bow shock, plasma conditions during T96 may not allow for the detection of ultra low frequency waves arising from the resonance between the IMF and exospheric pick-up particles.

In summary, Cassini observations during T96 have provided a unique look at the way in which Titan interacts with the supersonic solar wind. Because of the planned orbital geometry of the remaining Cassini Solstice Mission (with no future flybys on the dayside) it is certain that there will be not other chance to encounter Titan under these conditions during the Cassini era.

**References:**

Agren, K., et al., (2009). On the ionospheric structure of Titan. Planet. Space Sci., 57(14-15), 1821–1827. doi:10.1016/j.pss.2009.04.012.

Arridge, C. S., et al. (2011). Upstream of Saturn and Titan. Space Science Reviews, 162(1-4), 25–83. doi:10.1007/s11214-011-9849-x

Bertucci, C., Mazelle, C., & Acuña, M. (2005a). Structure and variability of the Martian magnetic pileup boundary and bow shock from MGS MAG/ER observations. Adv. Space Res., 36(11), 2066–2076. doi:10.1016/j.asr.2005.05.096.

Bertucci, C., Mazelle, C., Acuña, M. H., Russell, C. T., & Slavin, J. A. (2005b). Structure of the magnetic pileup boundary at Mars and Venus. J. Geophys. Res., 110(A1). doi:10.1029/2004JA010592.

Bertucci, C., et al. (2008). The Magnetic Memory of Titan's Ionized Atmosphere. Science, 321(5895), 1475–1478. doi:10.1126/science.1159780.

Bertucci, C., Duru, F., Edberg, N., Fraenz, M., Martinecz, C., Szego, K., & Vaisberg, O. (2011). The Induced Magnetospheres of Mars, Venus, and Titan. Space Sci. Rev., 162(1-4), 113–171. doi:10.1007/s11214-011-9845-1.

Coates, A. J., et al. (2012), Cassini in Titan's tail: CAPS observations of plasma escape, J. Geophys. Res., 117, A05324, doi:10.1029/2012JA017595.

Cravens, T. E., et al. (2010), Dynamical and magnetic field time constants for Titan's ionosphere: Empirical estimates and comparisons with Venus, J. Geophys. Res., 115, A08319, doi:10.1029/2009JA015050.

Dougherty, M., et al. (2004). The Cassini magnetic field investigation. Space Science Reviews, The Cassini-Huygens Mission, 331–383.

Dubinin, E., et al. (2006). Plasma Morphology at Mars. Aspera-3 Observations. Space Science Reviews, 126(1-4), 209–238. doi:10.1007/s11214-006-9039-4.

Edberg, N. J. T., Lester, M., Cowley, S. W. H., Eriksson, A. I. (2008). Statistical analysis of the location of the Martian magnetic pileup boundary and bow shock and the influence of crustal magnetic fields. J. Geophys. Res., 113(A8). doi:10.1029/2008JA013096.

Edberg, N. J. T., J.-E. Wahlund, K. Ågren, M. W. Morooka, R. Modolo, C. Bertucci, and M. K. Dougherty (2010), Electron density and temperature measurements in the cold plasma environment of Titan: Implications for atmo- spheric escape, Geophys. Res. Lett., 37, L20105, doi:10.1029/ 2010GL044544.

Garnier, P., Dandouras, I., Toublanc, D., Brandt, P. C., Roelof, E. C., Mitchell, D. G., Krimigis, S. M., et al. (2007). The exosphere of Titan and its interaction with the kronian magnetosphere: MIMI observations and modeling. Planetary and Space Science, 55(1-2), 165–173. doi:10.1016/j.pss.2006.07.006

Gurnett, D., et al. (2004). The Cassini radio and plasma wave investigation. Space Sci. Rev., 114(1), 395–463.

Krimigis, S., et al. (2004). Magnetosphere imaging instrument (MIMI) on the Cassini mission to Saturn/Titan. Space Science Reviews, The Cassini-Huygens Mission, 233–329.

Masters, A., N. Achilleos, M. K. Dougherty, J. A. Slavin, G. B. Hospodarsky, C. S. Arridge, and A. J. Coates (2008), An empirical model of Saturn's bow shock: Cassini observations of shock location and shape, J. Geophys. Res., 113, A10210, doi:10.1029/2008JA013276.

Ness, N. F., Acuña, M. H., Behannon, K. W., Neubauer, F. M. (1982). The induced magnetosphere of Titan. J. Geophys. Res., 87(A3), 1369–1381.

Neubauer, F. M., Backes, H., Dougherty, M. K., Wennmacher, A., Russell, C. T., Coates, A., Young, D., et al. (2006). Titan's near magnetotail from magnetic field and electron plasma observations and modeling: Cassini flybys TA, TB, and T3. J. Geophys Res., 111(A10). doi:10.1029/2006JA011676.

Romanelli, N., et al., Outflow and plasma acceleration in Titan's induced magnetotail: Evidence of magnetic tension forces, Submitted to J. Geophys. Res., 2014JA020391.

Sergis, N., Jackman, C. M., Masters, A., Krimigis, S. M., Thomsen, M. F., Hamilton, D. C., Mitchell, D. G., et al. (2013). Particle and magnetic field properties of the Saturnian magnetosheath: Presence and upstream escape of hot magnetospheric plasma. J. Geophys. Res. doi:10.1002/jgra.50164.

Sonnerup, B.U.O. and Scheible, M., Minimum and Maximum Variance Analysis in Analysis Methods for Multi-Spacecraft Data. Götz Paschmann and Patrick Daly (eds.), ISSI Scientific Reports Series SR-001, ESA/ISSI, Vol. 1. ISBN 1608-280X, 1998.

Thomsen, M. F. (2013). Saturn's magnetospheric dynamics. Geophys. Res. Lett., Saturn's Magnetospheric Dynamics, 40(20), 5337–5344. doi:10.1002/2013GL057967

Zhang, T. L., et al. (2008). Initial Venus Express magnetic field observations of the Venus bow shock location at solar minimum. Planetary and Space Science, 56(6), 785–789. doi:10.1016/j.pss.2007.09.012.

Schwartz, S.J., D. Burgess, W.P. Wilkinson, R.L. Kessel, M. Dunlop, H. Lühr, Observations of short large-amplitude magnetic structures at a quasi-parallel shock, J. Geophys. Res., Vol. 97, Issue A4, p. 4209–4227, 1992.

Delamere, P. A., & Bagenal, F. (2004). Pluto's kinetic interaction with the solar wind. Geophysical Research Letters, 31(4), 4807. doi:10.1029/2003GL018122

Treumann, R. A. (2009). Fundamentals of collisionless shocks for astrophysical application, 1. Non-relativistic shocks. The Astronomy and Astrophysics Review, 17(4), 409–535. doi:10.1007/s00159-009-0024-2.

Ulusen, D., Luhmann, J. G., Ma, Y. J., Mandt, K. E., Waite, J. H., Dougherty, M. K., Wahlund, J. E., et al. (2012). Comparisons of Cassini flybys of the Titan magnetospheric interaction with an MHD model: Evidence for organized behavior at high altitudes. Icarus, 217(1), 43–54. Elsevier Inc. doi:10.1016/j.icarus.2011.10.009

**Acknowledgments:** CB acknowledges the support from the Argentine National Agency for Promotion of Science and Technology (ANPCYT) through grant 2012-1763. The data for this study is available from the Planetary Data System (http://pds.jpl.nasa.gov)